\documentclass[sn-mathphys-ay, Namedate, iicol]{sn-jnl}% Math and Physical Sciences Author Year Reference Style
\usepackage{graphicx}    % Including figure files
\usepackage{amsmath}     % Advanced maths commands
\usepackage{amssymb}     % Extra maths symbols
\usepackage{amsfonts}    % Extra maths fonts
\usepackage{xspace}
\usepackage{amsthm}%
\usepackage{mathrsfs}%
\usepackage[title]{appendix}%
\usepackage{xcolor}%
\usepackage{textcomp}%
\usepackage{manyfoot}%
\usepackage{booktabs}%
\usepackage{listings}%
\newcommand{\Msun}{\,$M_{\odot}$\xspace}
\newcommand{\Rsun}{\,$R_{\odot}$\xspace}
\newcommand{\Msyr}{\,$M_{\odot}$\,yr$^{-1}$\xspace}
\newcommand{\kms}{\,km\,s$^{-1}$\xspace}

\newcommand{\ergs}{\,erg\,s$^{-1}$\xspace}
\newcommand{\gcm}{\,g\,cm$^{-1}$\xspace}
\newcommand{\gcmq}{\,g\,cm$^{-3}$\xspace}
\newcommand{\cmq}{\,cm$^{-3}$\xspace}
\newcommand{\pcms}{\,photons\,cm$^{-2}$\,s$^{-1}$\xspace}
\newcommand{\Ha}{H$\alpha$\xspace}
\newcommand{\Hb}{H$\beta$\xspace}
\newcommand{\Hg}{H$\gamma$\xspace}
\newcommand{\Hd}{H$\delta$\xspace}

\newcommand{\HeII}{He\,{\sc ii}\xspace}
\newcommand{\CIV}{C\,{\sc iv}\xspace}

\newcommand{\FeII}{Fe\,{\sc ii}\xspace}
\newcommand{\A}{\,\AA\xspace}
\newcommand{\apj}{Astrophys. J. } % Astrophysical Journal
\newcommand{\apjs}{Astrophys. J. Suppl. Ser. } % Astrophysical Journal Supplementary Series
\newcommand{\apss}{Astrophys. Space Sci. } % Astrophysics and Space Science
\newcommand{\aap}{Astron. Astrophys. } % Astronomy and Astrophysics
\newcommand{\mnras}{Mon. Not. R. Astron. Soc. } % Monthly Notices of the RAS
\begin{document}
%-------------------------------------------------------------------------------
\title[Type IIP SN~2024bch] 
{Type IIP SN~2024bch: Hydrodynamic model, shock breakout, and circumstellar
   interaction} 

\author*[1,2]{\fnm{V. P.} \sur{Utrobin}}\email{utrobin@itep.ru}

\author[2]{\fnm{N. N.} \sur{Chugai}}\email{nchugai@inasan.ru}


\affil*[1]{\orgname{NRC ``Kurchatov Institute''},
   \orgaddress{\street{acad. Kurchatov Square 1}, \city{Moscow},
   \postcode{123182}, \country{Russia}}}

\affil[2]{\orgname{Institute of Astronomy, Russian Academy of Sciences},
   \orgaddress{\street{Pyatnitskaya St. 48}, \city{Moscow},
   \postcode{119017}, \country{Russia}}}

\abstract{%
The well-observed type IIP SN~2024bch with the short plateau is shown to be
   an outcome of the red supergiant explosion with the presupernova mass of
   $14-15$\Msun, the explosion energy of $2\times10^{51}$\,erg, and presupernova
   radius of 1250\Rsun.
The early gamma-ray escape demonstrated by the radioactive tail is due to the
   large $^{56}$Ni extension up to 7400\kms.  
The early-time spectral evolution indicates the presence of the circumstellar
   dense confined envelope with the mass of $0.003-0.006$\Msun within
   $6\times10^{14}$\,cm.
The deceleration of the outermost ejecta implies the wind with the mass-loss
   rate of $\approx$6$\times10^{-4}$\Msyr.
The inferred mass-loss rate is by one-two order larger compared to most of
   type IIP supernovae, but comparable to the wind of type IIL SN~1998S.
The asymmetry of the broad \Ha component on day 144 powered by the circumstellar
   interaction is the outcome of the Thomson scattering and absorption
   in the Paschen continuum in the unshocked ejecta.
} 


\maketitle

%===============================================================================
\section{Introduction} 
\label{sec:intro}
%-------------------------------------------------------------------------------
Theory of stellar evolution predicts that type IIP supernovae (SNe~IIP) --- i.e.
   those that retain the massive hydrogen envelope prior to the core collapse ---
   originate from stars with the initial mass in the range of $9-25$\Msun
   \citep{Woosley_2002}.
The hydrodynamic modeling of well-observed SNe~IIP is able to determine the key
   parameters of SNe~IIP: the presupernova (pre-SN) mass, the explosion energy,
   the pre-SN radius, and the $^{56}$Ni mass with its extension in the ejecta.
Hereafter ``pre-SN'' means a star just before the explosion, while ``progenitor''
   means the star at the main sequence.

Generally, one expects that the growing number of SNe~IIP studied hydrodynamically
   in a uniform approach should recover the Salpeter initial mass function (IMF),
   with a small allowance for the mass lost via the stellar wind.
The Salpeter IMF predicts the number ratio ($\mathcal{R}$) of stars in the range
   of $9-15$\Msun to the number in the range of $15-25$\Msun to be
   $\mathcal{R} \approxeq 2$.

Surprisingly, the list of 14 SNe~IIP modeled hydrodynamically in a uniform way
   \citep{UC_2024} contains only two SNe~IIP in the range $M_{psn} < 15$\Msun
   with the resulting ratio $\mathcal{R} \sim 0.2$, being lower by a factor of
   ten.
The disparity requires an explanation that is still out.
It is noteworthy that the first evidence for the scarcity of the low-mass
   progenitors ($8-13$\Msun), based on the dissimilar mass estimates, was found
   by \cite{Chevalier_2006}, which led the authors to conclude that possibly not
   all stars in this mass range end up as SNe~IIP.

In this respect, of great interest is the recent hydrodynamic study of a large
   sample of SNe~IIb \citep{Ergon_2025} that recovers the progenitor masses
   based on the inferred mass of the He core in combination with the theoretical
   relation between the progenitor mass and the final helium core.
The study come up with the statement that the distribution of the SNe~IIb
   progenitor masses is consistent with the Salpeter IMF in the range of
   $9-25$\Msun.
The authors found a significant correlation between the helium-core mass and
   the explosion energy.
This means that low-mass massive stars explode in a normal way, although
   with the lower energy than more massive stars.

The unsettled issue of the mass distribution of the SNe~IIP progenitors is
   a primary motivation to increase the sample of the well-observed SNe~IIP
   studied hydrodynamically in a uniform approach.
The first attempts to evaluate the three basic physical parameters (the initial
   radius, the ejecta mass, and the explosion energy) from the values measured
   from observations (the characteristic duration of the light curve,
   the bolometric luminosity and the photospheric velocity in the middle of
   the plateau) were carried out by \citet{LN_1983, LN_1985}.
Unfortunately, this approach contains the internal inconsistency: it ignores
   completely the significant contribution of the initial luminosity peak to
   the derived parameters.
This shortcoming results in the degeneracy of SN~IIP parameters recovered from
   the bolometric luminosity and the photospheric velocity in the middle of
   the plateau \citep{Goldberg_2019}.
To overcome this problem and to construct an adequate hydrodynamic model of
   SN~IIP, not only the bolometric light curve as a whole or the photometric
   observations, but also the expansion velocities, especially at the early
   stage ($t < 40$ days), should be taken into account.

The well-observed SN~2024bch \citep{Andrews_2025} is a highly interesting case
    because of its unconventional observational properties.
It demonstrates particularly: (i) a short (70 days) steeply decaying plateau;
   (ii) a signature of the early escape of gamma-quanta at the radioactive tail;
   (iii) an intermediate width (1000\kms) \Ha absorption with the peculiar
   profile against the featureless continuum on day 8;
   (iv) a boxy \Ha emission on day 144 indicating the ejecta interaction with
   a dense wind.
The first and second properties could be caused by the low-mass ejecta,
   although they could be the outcome, as well, of an appropriate combination
   of the SN parameters with the massive ejecta likewise SN~2018gj \citep{UC_2024}.
The third and fourth properties make SN~2024bch similar to type IIL SN~1998S
   \citep{Fassia_2001}; the spectral similarity of these SNe is emphasized
   by \cite{Andrews_2025}.

We start, therefore, with general remarks on the relation between the early
   spectral evolution of these two supernovae and background physics that can
   become the useful landmarks for hydrodynamic study (Section~\ref{sec:genphys}).
We then describe the pre-SN model and the hydrodynamic modeling in comparison
   with observations of SN~2024bch, including the early circumstellar (CS)
   interaction and the preshock acceleration seen in the \Ha profile on day 8
   (Section~\ref{sec:hydro}).
In Section~\ref{sec:psnwind} we explore the issue of the late boxy \Ha emission
   and recover the wind density based on the deceleration effects of the outermost
   ejecta expanding in the circumstellar matter (CSM).
Finally, in Section~\ref{sec:disc} we summarize and discuss our results.

%===============================================================================
\section{Physics behind early spectra}
\label{sec:genphys}
%-------------------------------------------------------------------------------
The spectral evolution of SN~2024bch during the first month after the explosion
   \citep{Tartaglia_2024, Andrews_2025} in many ways is similar to that of
   type IIL SN~1998S \citep{Fassia_2001}.
Both supernovae evolve through three major stages
   with a clear physical meaning \citep{Chugai_2001}.

At the first stage ($t < 5$ days), the SN~2024bch spectrum shows the CS
   emission lines of \Ha and high-ionization lines of \HeII 4686\,\A and
   \CIV 5805\,\A on the top of a smooth continuum.
Their specific profile --- narrow core and broad wings --- indicates that
   the SN ejecta expand through the highly ionized CS dense confined shell
   (DCS) with the Thomson optical depth $\tau_{\mbox{\tiny T}} \sim 2$ likewise
   in early SN~2013fs \citep{Chugai_2020}.
The broad wings form due to the multiple scattering of narrow CS emission lines
   off thermal electrons of the DCS with the temperature of $\gtrsim 20\,000$\,K,
   although a high velocity of the preshock gas accelerated by the SN radiation 
   can contribute to broadening as well \citep{Chugai_2020}. 

The disappearance of these emission lines after day 5 signals that the SN ejecta
   with the boundary velocity $v_{sn}\sim10^4$\kms emerge from the DCS with
   $\tau_{\mbox{\tiny T}} \sim 2$,
   the outer radius $R \approx v_{sn}t \sim 5\times10^{14}$\,cm, the number
   density $n \sim \tau_{\mbox{\tiny T}}/(R\sigma_{\mbox{\tiny T}}) \sim 6\times10^9$\cmq, and
   the DCS mass of $(4\pi/3)R^3nm_p \sim 3\times10^{-3}$\Msun.
It is remarkable that the inferred DCS parameters are similar to those of
   type IIP SN~2013fs with the well-observed ionization flash spectrum
   \citep{Yaron_2017}.
At this stage, within time span of 10 days, the featureless continuum indicates
   that the photosphere resides at the boundary opaque cold dense shell (CDS)
   which conceals the unshocked ejecta.
This explains the absence of broad P Cygni lines.

The only line seen at this stage ($t = 8$ days) is the conspicuous \Ha absorption
   with the width of $\approx$1000\kms and the maximum depth close to zero
   velocity \citep{Andrews_2025}.
This absorption apparently forms in the accelerated preshock CSM likewise
   a similar \Ha absorption in early SN~1998S \citep{Chugai_2002}.
This issue is discussed in more detail below (Section \ref{sec:csint}).

Broad P Cygni lines, characteristic of SNe~IIP, emerge at the third stage,
   after day 15.
This fixes the moment when the CDS becomes transparent and the photosphere
   recedes into the unshocked ejecta.
Note that in the case of SN~1998S the CDS becomes transparent at about 40 days
   after the explosion, which suggests a more massive CDS compared to SN~2024bch
   by at least a factor of $\sim$(40/15)$^2 = 7$.
The transition moments from the first stage to the second stage ($t_1\approx5$ days)
   and from the second stage to the third stage ($t_2\approx15$ days) are landmarks
   for the hydrodynamic model and the CSM distribution in the close vicinity of
   the pre-SN star.

%===============================================================================
\section{Hydrodynamic modeling}
\label{sec:hydro}
%-------------------------------------------------------------------------------
%
%===============================================================================
\subsection{Supernova parameters}
\label{sec:snpar}
%-------------------------------------------------------------------------------
%
\begin{figure}
   \includegraphics[width=\columnwidth, clip, trim=0 238 52 139]{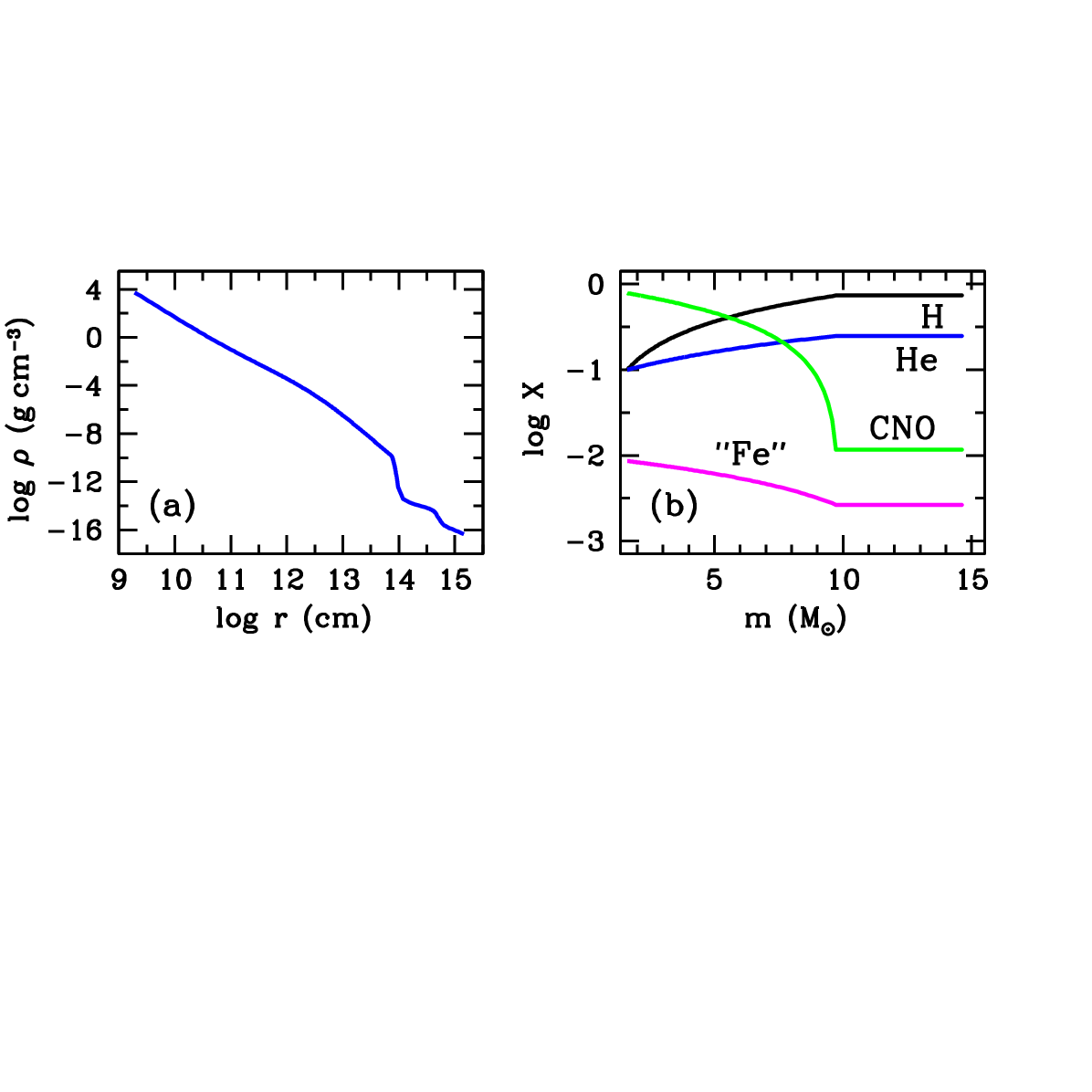}
   \caption{%
   The structure of the pre-SN model.
   Panel (a): the density distribution as a function of radius.
   At the radii $r > 10^{14}$\,cm the density refers to CSM.
   Panel (b): the chemical composition.
   Mass fraction of hydrogen (\emph{black line\/}), helium
      (\emph{blue line\/}), CNO elements (\emph{green line\/}),
      and Fe-peak elements excluding radioactive $^{56}$Ni
      (\emph{magenta line\/}) in the ejected envelope.
   The central core of 1.6\Msun is omitted.
   }
   \label{fig:presn}
\end{figure}
The modeling of SN~2024bch is performed using the Lagrangian radiation
   hydrodynamics (RHD) code {\sc CRAB} \citep{Utrobin_2004, Utrobin_2007}
   modified by introducing the artificial acceleration of mixing
   to treat large density contrasts developing in the one-dimensional (1D) modeling
   of shock propagation in radiating fluids similar to \citet{Blinnikov_1998}.

The pre-SN is the hydrostatic nonevolutionary red supergiant (RSG) model with
   the artificial mixing that mimics three-dimensional (3D) mixing during
   the shock wave propagation in the exploding star \citep{UWJM_2017}.
The initial configuration with the adopted density and composition distributions
   is presented in Fig.~\ref{fig:presn}.
The density distribution includes the CSM consisting of the DCS and the external
   steady RSG wind.
The CSM structure is constrained primarily by the early spectra evolution.

\begin{figure}
   \includegraphics[width=\columnwidth, clip, trim=0 239  53 134]{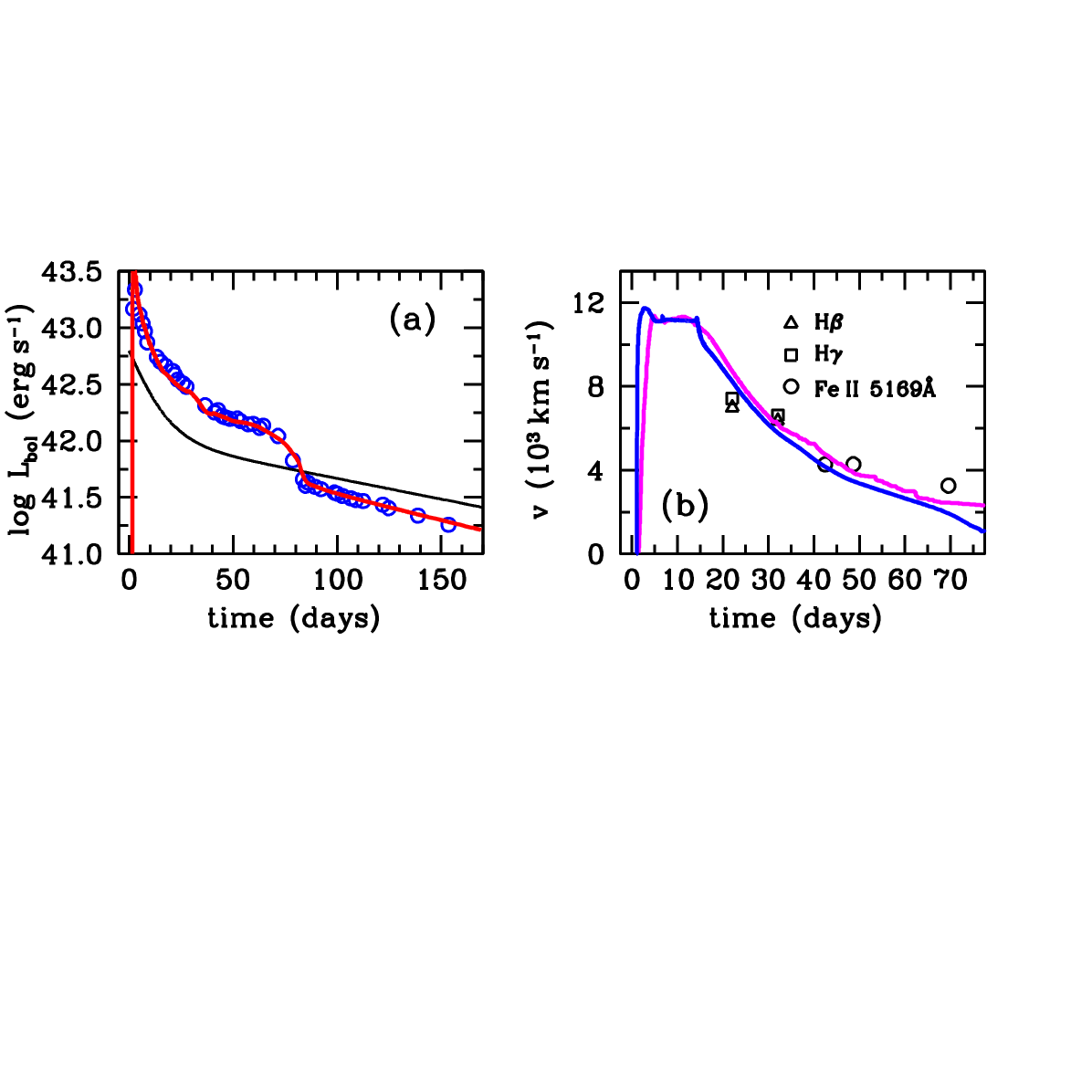}
   \caption{%
   The bolometric light curve and the evolution of photospheric velocity.
   Panel (a): the model light curve (\emph{red line\/}) overlaid on
      the bolometric data (\emph{circles\/}) \citep{Andrews_2025}.
   The \emph{black line} is the total power of radioactive $^{56}$Ni decay.
   Panel (b): the evolution of model velocity defined by the level
      $\tau_{eff} = 2/3$ (\emph{blue line\/}) and $\tau_\mathrm{Thomson} = 1$
      (\emph{magenta line\/}) is compared with the photospheric velocities
      estimated from the absorption minimum of \FeII 5169\A \citep{Tartaglia_2024}
      along with our estimates from the \Hb and \Hg lines \citep{Andrews_2025}.
   }
   \label{fig:lcv}
\end{figure}
The explosion is initiated by a supersonic piston at the boundary with
   the 1.6\Msun collapsing core.
The optimal description of the bolometric light curve and the expansion
   velocities (Fig.~\ref{fig:lcv}) is found for the pre-SN radius $R_0$, the ejecta
   mass $M_{ej}$, the explosion energy $E$, the $^{56}$Ni mass $M_{\mathrm{Ni}}$
   and its mixing extension $v_{\mathrm{Ni}}^{max}$ shown
   in Table~\ref{tab:param}.
A difference between the bolometric light curves obtained by \citet{Andrews_2025}
   and \citet{Tartaglia_2024} is used as a measure to estimate the parameter
   errors listed in Table~\ref{tab:param}.
The effects of parameter variation on the SN~IIP model were studied in detail
   and illustrated before \citep{Utrobin_2007}.

The bolometric maximum of the hydrodynamic model occurs on day 1.9 after
   the SN explosion.
As the forward shock propagates and enters the DCS, the photospheric velocity
   decreases starting from day 2.6 (Figs.~\ref{fig:lcv}b and \ref{fig:nocsm}c).
Around day 6 the shock leaves the DCS and enters the RSG wind.
Next 9 days the photospheric velocity remains constant, manifesting
   the fact that the photosphere resides at the boundary opaque CDS.
The above behavior of the optimal model is in a good agreement with the first
   and second stages described in Section~\ref{sec:genphys}.

Note that the DCS mass and its extension used in the hydrodynamic model are
   not unique. 
The reported value of 0.0032\Msun should be considered as a minimum mass that
   is consistent with the duration of the opaque CDS stage $t_2\approx15$ days.
For the larger DCS mass, the CDS velocity and the DCS radius should be somewhat
   smaller to be compatible with the duration of phases $t_1\approx5$ days
   and $t_2\approx15$ days (Section~\ref{sec:genphys}).
  
Remarkably, the model shows a curious behavior of a ``thomsonsphere'' (i.e.,
   the level of $\tau_{\mbox{\tiny T}}=1$) during the initial velocity rise to
   the maximum value (Fig.~\ref{fig:lcv}b): the thomsonsphere velocity at this
   stage turns out significantly lower compared to the velocity
   at the photosphere.
The reason is that the thomsonsphere forms in the accelerated CSM, while
   the photosphere resides at the effective optical depth $\tau_{eff}=2/3$
   in the SN ejecta with the larger expansion velocity.

It should be emphasized that the early escape of gamma-quanta indicated by
   the steep radioactive tail of SN~2024bch is related not to a low ejecta
   mass but rather to the extended $^{56}$Ni distribution.
In addition, the plateau of the bolometric light curve of SN~2024bch with
   the relatively short duration and the rapid luminosity decline is
   consistent with a relatively large ratio $E/M_{ej}$ and an extended
   $^{56}$Ni distribution.

To study a sensitivity of the bolometric light curve to the degree of the
   $^{56}$Ni mixing, we calculate a special model which is identical to
   the optimal model, but has the $^{56}$Ni mass of 0.06\Msun and 
   the moderate $^{56}$Ni mixing within 4000\kms.
It is interesting that this model differs from the optimal model only
   in its radioactive tail which does not show any signature of the
   gamma-quanta escape.

\begin{figure}
   \includegraphics[width=\columnwidth, clip, trim=0 239 53 139]{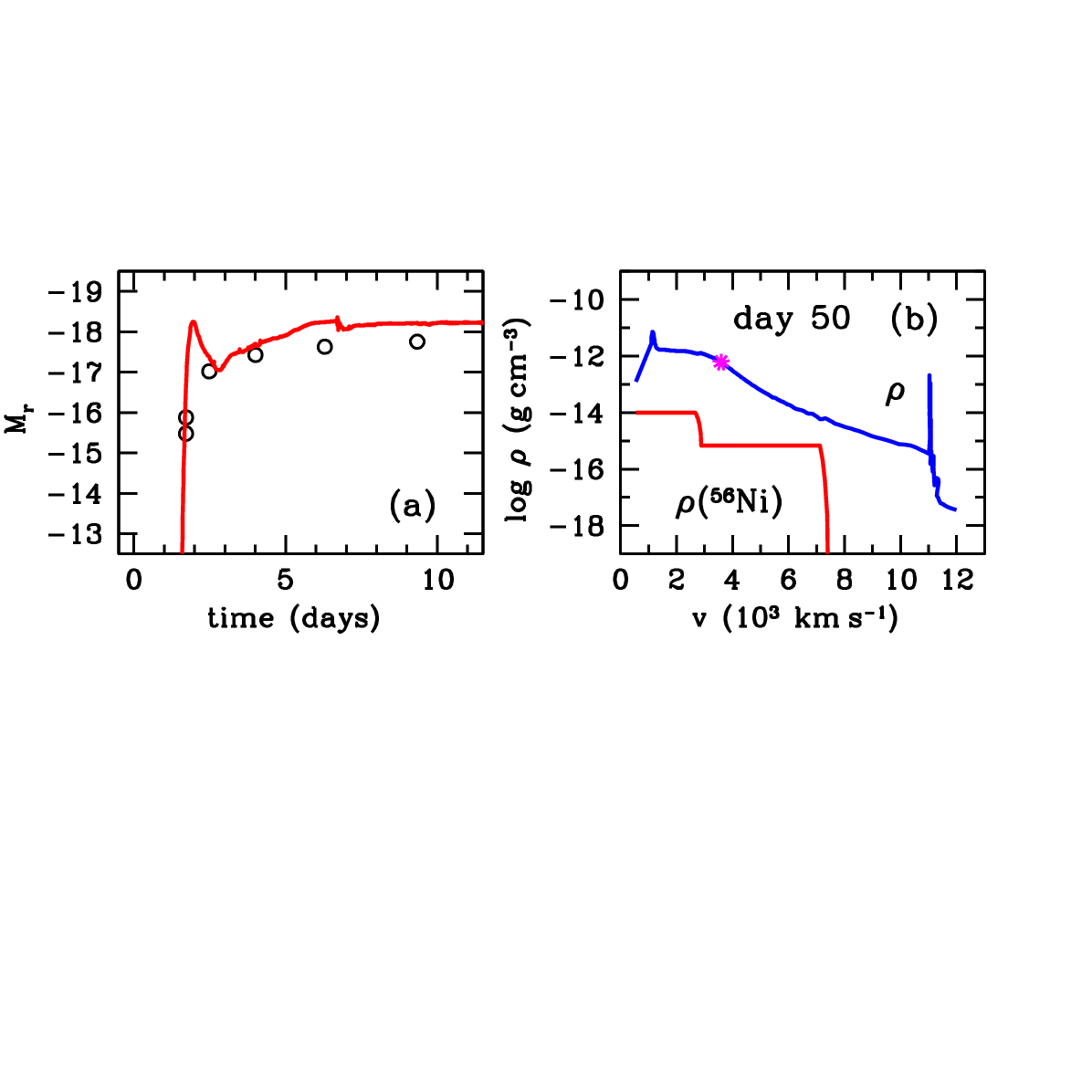}
   \caption{%
   Panel (a): Rising part of the model light curve in the $r$-band overplotted
      on the observational data taken by \citet{Tartaglia_2024}.
   Panel (b): The density and $^{56}$Ni distributions vs. velocity in the ejecta
      on day 50; magenta star indicates the photosphere location.
   }
   \label{fig:rise}
\end{figure}
\begin{table}
\centering 
\caption{Parameters of the optimal model}
\label{tab:param}
\begin{tabular}{@{ } l @{ } c @{ } c @{ } c @{ }}
\toprule
\noalign{\smallskip}
Parameter & Unit & Value & Error \\
\noalign{\smallskip}
\midrule
\noalign{\smallskip}
Pre-SN radius              & \Rsun	         & 1250               & $\pm$\,200   \\
Ejected mass               & \Msun          & 13.0               & $\pm$\,3.0   \\
Explosion energy           & $10^{51}$\,erg & 2.0                & $\pm$\,0.36  \\
$^{56}$Ni mass             & \Msun          & 0.075              & $\pm$\,0.007 \\
Extent of $^{56}$Ni mixing & \kms           & 7400               & $\pm$\,200   \\
DCS mass                   & \Msun          & $3.2\times10^{-3}$ & --           \\
\botrule
\end{tabular}
\end{table}
The rising part of the $r$-band light curve (Fig.~\ref{fig:rise}a) fixes the
   explosion moment at JD\,2460337.85, i.e., 0.2 days earlier compared to that of
   \citet{Andrews_2025}.
The total density and the $^{56}$Ni density in the freely expanding ejecta
   are shown in Fig.~\ref{fig:rise}b on day 50.
The pronounced boundary density peak is the CDS that becomes transparent
   in continuum after day 15 in line with the spectral signatures of SN~2024bch
   (see Section~\ref{sec:genphys}).
Generally, the decelerating CDS is subject to the Rayleigh-Taylor instability that, however,
   cannot be simulated by 1D hydrodynamics.  

%===============================================================================
\subsection{Early CS interaction}
\label{sec:csint}
%-------------------------------------------------------------------------------
After the RSG explosion and the shock breakout (SBO), the SN in the empty
   environment would look like an expanding ``fireball'' bounded by
   a low-mass stellar CDS \citep{Chevalier_1976, GIN_1971}.
In contrast, in the dense CSM the outer layers of the SN ejecta are decelerated
   via the forward and reverse shocks with the swept-up shocked CSM
   and the ejecta accumulated at the contact surface \citep{Chevalier_1982b}.
This is a standard picture of the SN/CSM interaction at the relatively late stage,
   when the role of the radiation-driven flow becomes negligible.

At the early stage ($t < 10$ days), however, this picture is substantially
   modified by the powerful SN radiation.
The radiative acceleration of the CS preshock gas may turn out so large that
   the viscous jump almost disappears, which severely suppresses the forward
   post-shock gas temperature.
This effect is of importance for the interpretation of early hard X-rays
   from SNe in the dense CS environment.  
 
%===============================================================================
\subsubsection{CSM effects in the RHD model} 
\label{sec:impact}
%-------------------------------------------------------------------------------
%
\begin{figure}
   \includegraphics[width=\columnwidth, clip, trim=0 18 54 134]{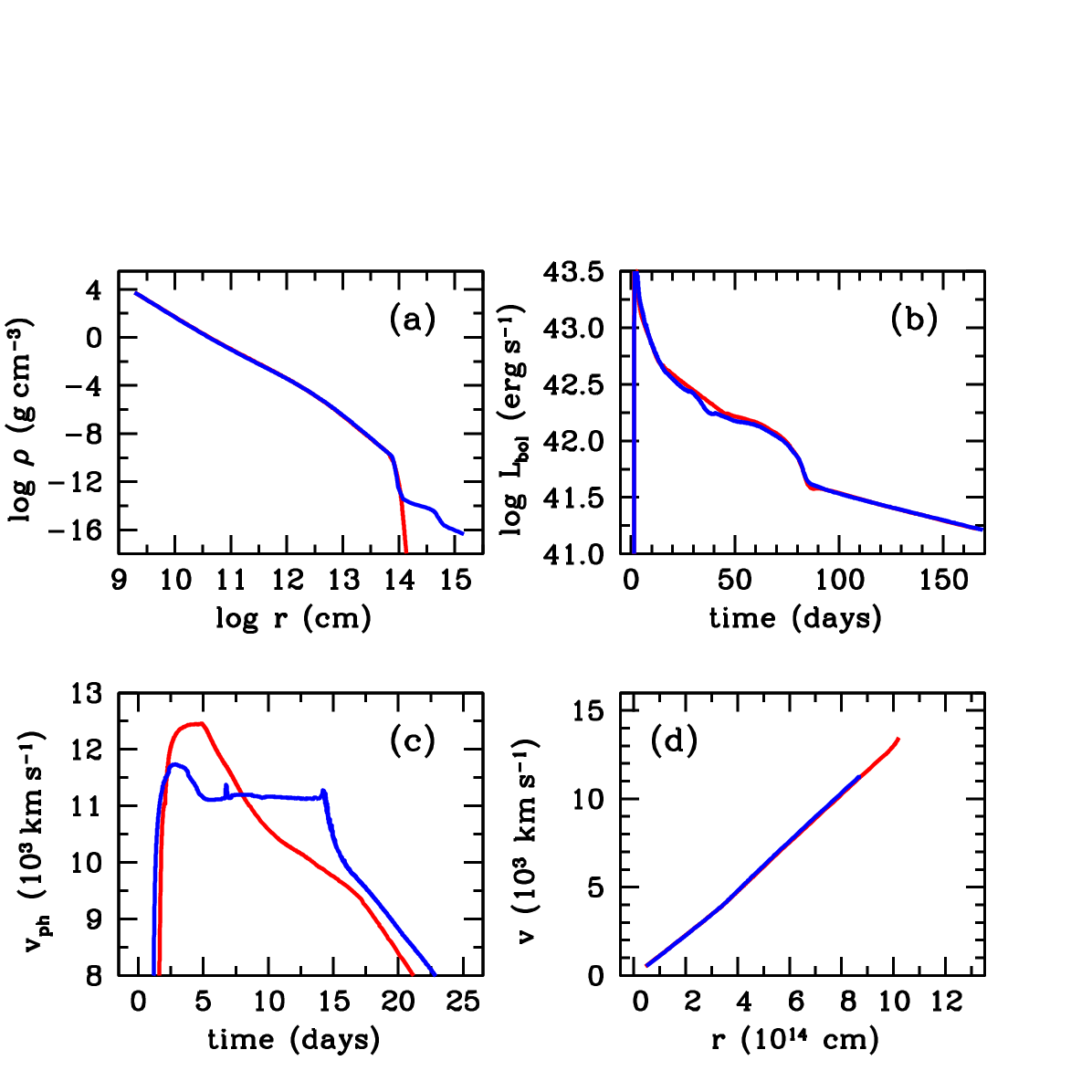}
   \caption{%
   Effects of the CSM on the light curve and the ejecta velocity.
   The \emph{blue line} is the optimal model with the CSM and
      the \emph{red line} is the model without the CSM.
   Panel (a): the density distribution for both pre-SN models;
   Panel (b): the corresponding bolometric light curves;
   Panel (c): the velocity at the photosphere as a function of time;
   Panel (d): the velocity distribution in the unshocked SN ejecta on day 10.
   The effect of the CSM on the light curve is negligible, but pronounced on
      the velocity of the outer layers.
   }
   \label{fig:nocsm}
\end{figure}
The effects of the CSM on the light curve and the expansion velocities are
   demonstrated by the optimal model with the CSM compared to the
   model without the CSM (Fig.~\ref{fig:nocsm}).
The basic parameters of these models are very close and differ only by a factor
    less than 4\%.
The density distributions of both pre-SN models are shown in
   Fig.~\ref{fig:nocsm}a.
The DCS mass in the optimal model is too small ($\sim$0.003\Msun) to noticeably
   affect its light curve, so the light curves of both models are essentially
   similar with a small difference around day 40 (Fig.~\ref{fig:nocsm}b).

The pronounced difference, however, is revealed by the photospheric velocities
   between days 5 and 15, when the optimal model demonstrates the constant
   velocity stage with the photosphere residing at the boundary opaque CDS
   (Fig.~\ref{fig:nocsm}c).
After day 15 the CDS becomes transparent and the photosphere starts to recede
   into the unshocked ejecta, which is marked by the appearance of broad P Cygni
   lines at $t > 15$ days.
Finally, in the model without the CSM the radial distribution of velocity
   in the unshocked ejecta on day 10 exhibits higher outer velocities
   compared to those of the optimal model (Fig.~\ref{fig:nocsm}d).
This difference could be revealed by a higher velocity of
   the absorption blue wing of hydrogen Balmer lines in early spectra 
   for the model either without, or with, a rarefied CSM.

%===============================================================================
\subsubsection{From SBO to adiabatic forward shock} 
\label{sec:cdsreg}
%-------------------------------------------------------------------------------
%
\begin{figure}
   \includegraphics[width=\columnwidth, clip, trim=0 10 27 28]{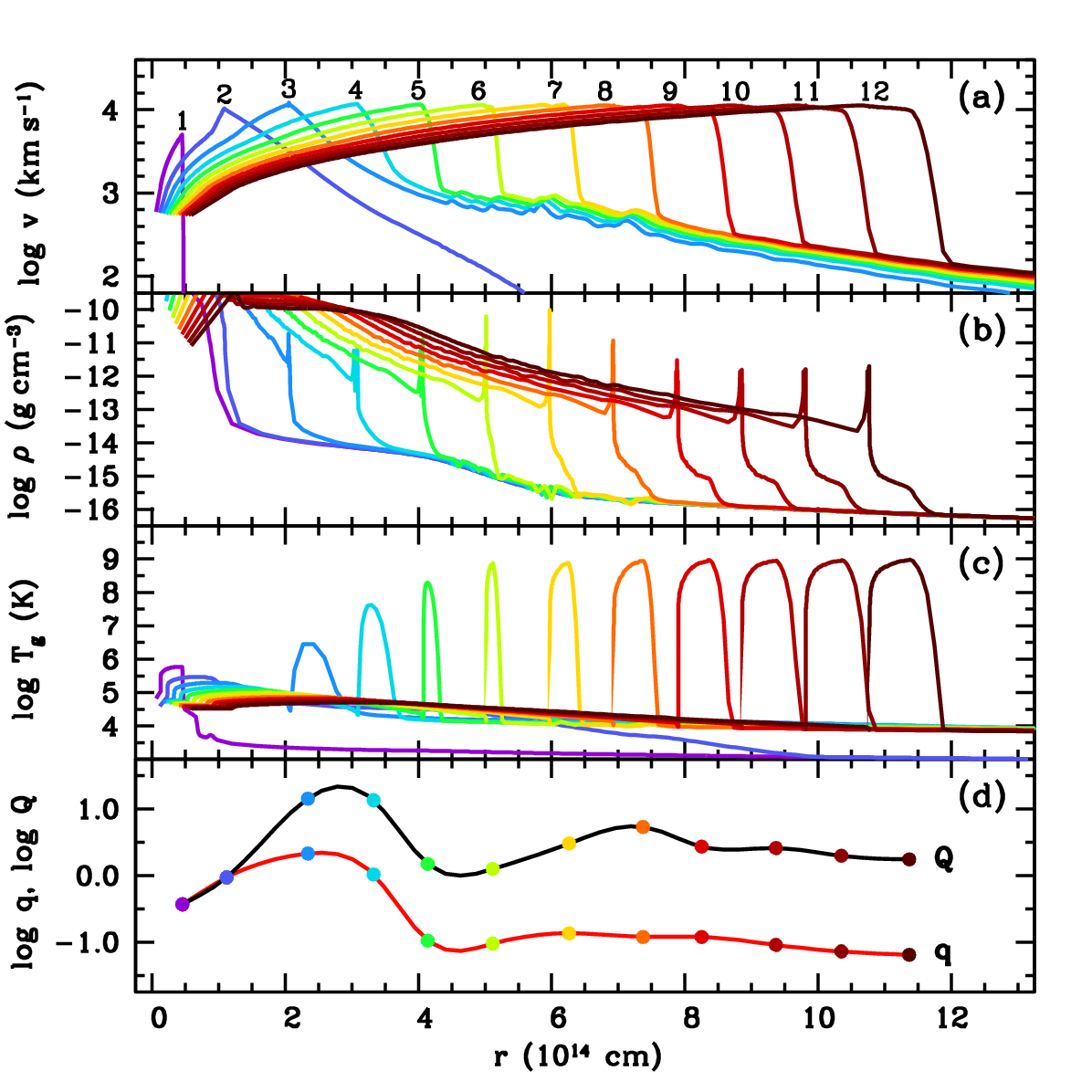}
   \caption{%
   The evolution of the radial profiles of velocity (Panel (a)),
      density (Panel (b)), and gas temperature (Panel (c)) from day 1 to
      day 12 shows the forward shock structure and the radiative acceleration
      of the preshock CS gas.
   Profiles are plotted at the elapsed times in days indicated in Panel (a).
   Panel (d) shows the parameters $q$ (\emph{red line}\/) and $Q$
      (\emph{black line}\/); colored circles correspond to the moments
      pointed out in Panel (a).
   }
   \label{fig:shock}
\end{figure}
The evolution of the radiative hydrodynamic flow\footnote{%
   Note that the fronts of the reverse and forward shocks are smeared due to
      the artificial viscosity on the computational Lagrangian mass grid with
      a rather crude spatial resolution, especially in the rarefied CSM.}
   between days 1 and 12 is illustrated by the radial profiles of velocity,
   density, and gas temperature in Figs.~\ref{fig:shock}a, b, and c.
In Fig.~\ref{fig:shock}d we additionally show two relevant parameters $q$
   and $Q$:
$$
   q = \frac{\Delta L_{shk}}{L_{kin}}, \; \;
   Q = \frac{L_{shk}}{L_{kin}}
             \qquad \mbox{for} \; \; t > 2\;\mbox{days} \; ,
$$
where $\Delta L_{shk}$ is the net luminosity produced by the gas between
   the CDS and the forward shock;
   $L_{shk}$ is the total shock luminosity (outgoing radiation of the ejecta
   plus radiation generated by the forward shock);
   $L_{kin}$ is the kinetic energy luminosity of the inflowing gas
   at the forward shock.
Note that the parameters $q$ and $Q$ are equal for $t \le 2$ days.

The behavior of $Q(t)$ emphasizes the significant role of the SN radiation
   ($Q \gg q$), whereas $q(t)$ demonstrates the transition of the forward shock
   towards the adiabatic regime with the high post-shock gas temperature of
   $\sim$10$^9$\,K.

On day 1 the radiation-dominated shock propagates through the outer layers of
   the pre-SN (Figs.~\ref{fig:shock}a and b).
During the SBO the bolometric luminosity attains its maximum at about day 2
   and the radiation begins to accelerate the CSM (Fig.~\ref{fig:shock}a).
It should be emphasized that the CSM optical depth ($\tau \sim 2-3$) is
   significantly lower than the critical value $c/v \approx 30$ 
   (with $v\approx10^4$\kms) required for the radiation trapping,
   so the CSM does not affect the SBO.
The strong radiation flux maintains the isothermal shock regime for 
   a couple of days and by day 4 almost all the DCS is swept up into
   the opaque CDS (Fig.~\ref{fig:shock}b).
Later on, the CDS operates as a piston and drives the matter-dominated shock.

After day 3 the gas temperature behind the shock decouples from the radiation
   temperature of $\sim$($1-3$)$\times10^4$\,K and rises towards the value of
   $\sim$10$^9$\,K prescribed by the Rankine-Hugoniot jump conditions
   (Fig.~\ref{fig:shock}c).
Note that between days 2 and 5 the model post-shock gas temperature is probably
   overestimated, since code {\sc CRAB} does not include the net line emission
   that dominates the cooling rate in the temperature range of
   $10^4 - 3\times10^7$\,K \citep{Sutherland_1993}.

\begin{figure}
   \includegraphics[width=\columnwidth, clip, trim=19 113 20 116]{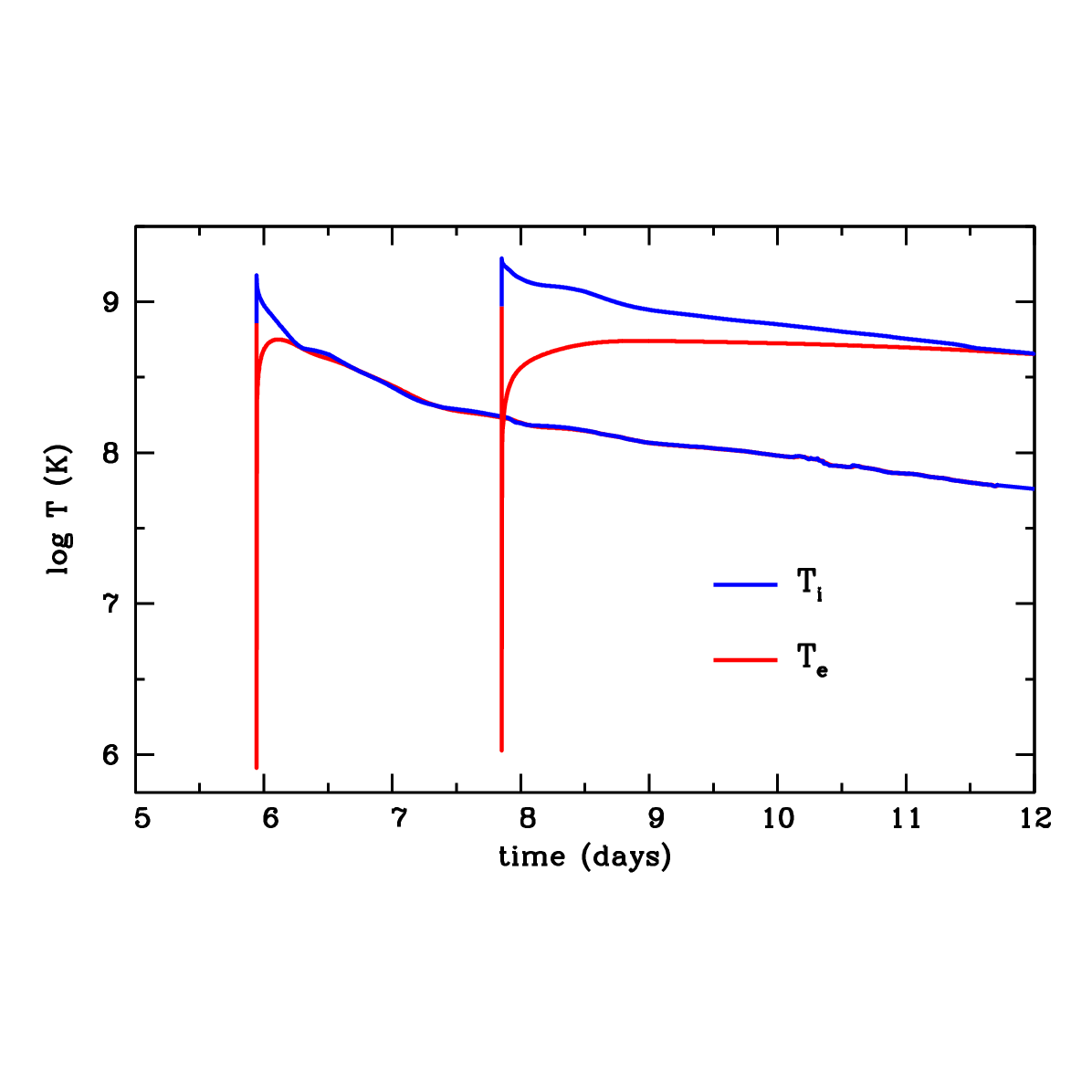}
   \caption{%
   The post-shock equilibration of electron (\emph{red line\/}) and ion
      (\emph{blue line\/}) temperatures in two Lagrangian mass zones of
      the optimal model.
   The shock subsequently passes these zones in the CSM at around days 6
      and 8, respectively.
   }
   \label{fig:tauei}
\end{figure}
The ion and electron temperatures ($T_i$ and $T_e$) at the forward shock front
   are determined by the Rankine-Hugoniot jump conditions for each component
   in proportion to their masses, i.e., $T_e \approx (m_e/m_p)T_i$.
The post-shock temperature equilibration proceeds due to the Coulomb $e$-$i$
   collisions \citep{Spitzer_1962} with a minor contribution of the Compton
   cooling \citep{Weymann_1966} of hot electrons in the SN radiation field.

In the optimal model, the forward shock crosses two representative Lagrangian
   mass zones at around days 6 and 8 and the following post-shock temperature
   evolution shows that the electron temperature rises at the time scale of
   about 0.2 days and 0.5 days, respectively (Fig.~\ref{fig:tauei}).
The electron temperature at this stage, therefore, has enough time to rise
   up to $\sim$5$\times10^8$\,K.

With that high electron temperature SN~2024bch might have been a source of
   hard X-rays at the early stage $t > 5$ days.
For the adopted wind density the hard X-ray luminosity on day 8 presumably is
   $L_{\mbox{\tiny X}} \approx 2\times10^{41}$\ergs with the temperature of
   $kT \approx 40$\,keV.

%===============================================================================
\subsubsection{Preshock acceleration} 
\label{sec:radacc}
%-------------------------------------------------------------------------------
We now focus at the origin  of the \Ha absorption on day 8 that presumably forms
   in the preshock gas accelerated up to $u \approx 1000$\kms.
The effect of the radiative preshock CSM acceleration in the optimal model is
   rather pronounced during the first 10 days after the explosion
   (Fig.~\ref{fig:shock}a).
However, on day 8 the model preshock velocity is only 500\kms, twice as low
   compared to the value of 1000\kms that is suggested by the observed \Ha
   absorption \citep{Andrews_2025}.
The velocity mismatch implies that either the hydrodynamic model underestimates
   the expansion opacity of lines in the preshock gas, or an additional
   acceleration could be provided by the cosmic ray (CR) precursor.

The radiative acceleration could be underestimated, because the one-group
   radiation transfer cannot describe the additional heating of the CSM
   by the X-rays from the forward shock.
The effect of the gas temperature on the expansion line opacity can be
   demonstrated for the wind density of $10^{-15}$\gcmq and the velocity
   gradient $u/r \approx 8\times10^6$\,s$^{-1}$.
In this case the ratio of the line-to-Thomson opacities is
   $\kappa_{\mbox{\tiny L}}/\kappa_{\mbox{\tiny T}} = 3.6$ for
   the model with $T_e = 2\times10^4$\,K, whereas for $T_e = 4\times10^4$\,K
   the ratio is $\kappa_{\mbox{\tiny L}}/\kappa_{\mbox{\tiny T}} =  15$,
   i.e., four times larger.

Alternatively, the additional acceleration might be produced by the CR precursor.
The efficiency of the CR diffusive shock acceleration (DSA)
   \citep{Krymskii_1977,Bell_2004} defined as the ratio $\eta$ of relativistic
   pressure $P_r$ (CR plus magnetic field) to the upstream dynamic pressure
   $\rho_0v_0^2$ is equal approximately to the ratio of the additional preshock
   velocity produced by the relativistic precursor ($\Delta u$) to the shock
   velocity,  i.e., $\eta \approx \Delta u/v_0$ \citep{Chugai_2021}.
For SN~2024bch on day 8, with $\Delta u \approx 500$\kms and
   $v_0 \approx 10^4$\kms, the required efficiency should be $\eta \approx 0.05$.
That large CR acceleration efficiency on day 8 is plausible, but requires the
   independent confirmation. 
Thus, at the moment the origin of the high velocity of the preshock gas
   on day 8 remains an open issue.

%===============================================================================
\section{Presupernova wind}
\label{sec:psnwind}
%-------------------------------------------------------------------------------
%
%===============================================================================
\subsection{Asymmetric boxy \Ha on day 144}
\label{sec:boxyha}
%-------------------------------------------------------------------------------
%
\begin{figure}
   \includegraphics[width=\columnwidth, clip, trim=13 113  27 116]{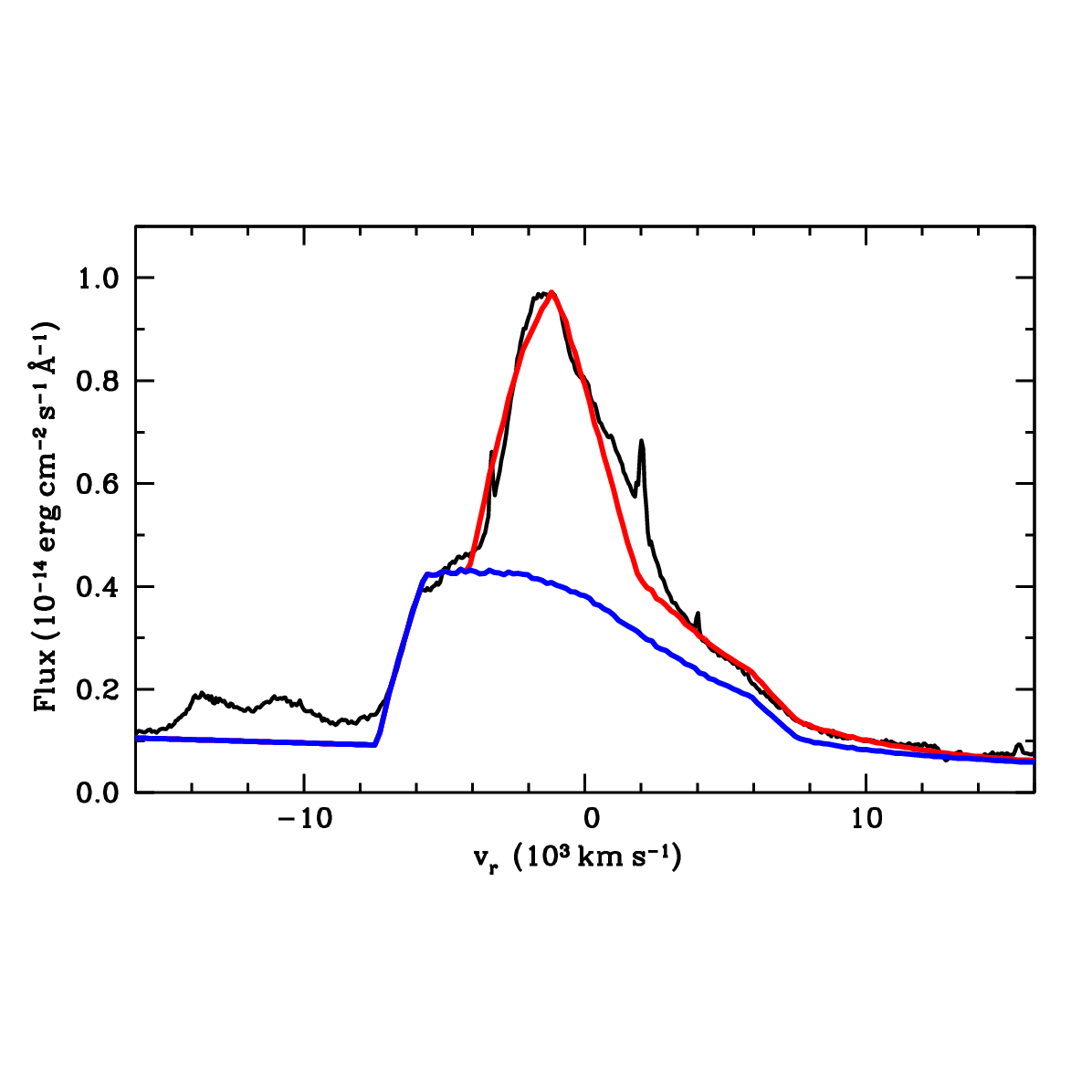}
   \caption{%
   \Ha emission on day 144 \citep{Andrews_2025} (\emph{black line\/})
      compared to the model profile that is a combination of the spherical
      boxy component (\emph{blue line\/}) and the spherical blueshifted central
      narrow component (\emph{red line\/}).
   Both components are modified by the Thompson scattering and the absorption
      in the Paschen continuum.
   The broad boxy component is powered primarily by the CS interaction,
      whereas the narrow component is presumably energized by the $^{56}$Co decay,
      which indicates the anisotropy of the $^{56}$Ni ejecta.
   }
   \label{fig:habox}
\end{figure}
After two weeks the SN expands in the slow RSG wind.
The wind density is moderate, since the CS interaction does not affect
   the bolometric luminosity, but the wind is dense enough to noticeably
   decelerate the outer layers of the ejecta on the time scale of 100 days.
Indeed, the spectrum on day 32 \citep{Andrews_2025} shows a clear-cut \Hd
   absorption with the blue edge at the velocity of the unshocked ejecta
   $v_{sn} = 8500\pm500$\kms.
On the other hand, on day 144 the blue edge of the boxy profile of the \Ha
   emission indicates the maximum ejecta velocity of $v_{sn} = 7500\pm500$\kms.
The latter value assumes the spherical wind geometry, whereas this might not be
   the case as suggested earlier \citep{Andrews_2025} given the \Ha blueshift.
The wind asphericity is a radical conjecture and, therefore, one needs to examine
   a natural possibility that the \Ha blueshift on day 144 is due to the Thomson
   scattering in combination with the Paschen continuum absorption
   in the spherical case.

We consider a homogeneous homologously expanding sphere with the boundary
   velocity  $v_{sn} = 7500$\kms.
The line-emitting region presumably consists of the outer zone
   $v_1 < v < v_2 = v_{sn}$ with $v_1 = 5800$\kms and the inner zone
   $v \leq 3000$\kms responsible for the central narrow component.
The inner zone is presumably powered by the radioactivity, whereas
   the outer zone is powered mainly by the X-ray radiation from
   the reverse and forward shocks.
We assume a homogeneous distribution of free electrons with the Thomson
   optical depth $\tau_{\mbox{\tiny T}}$ and the uniform Paschen continuum
   absorbers with the optical depth $\tau_a$.

Monte Carlo simulations of \Ha on day 144 (Fig.~\ref{fig:habox}) reproduce
   the asymmetric profile for the extinction optical depth $\tau_{ext} = 1$
   with $\tau_{\mbox{\tiny T}} = 0.6$ and $\tau_a = 0.4$;
   the uncertainty of these values is 15\%.
The fraction of the escaped \Ha radiation in this case turns out $\phi = 0.71$.
We apply a blue shift $v_{bs} = -1200$\kms for the inner \Ha-emitting zone
   to match the observed shift of the narrow component \citep{Andrews_2025},
   probably related to the anisotropy of the $^{56}$Ni ejecta.

Comparable values inferred above for the scattering and absorption optical
   depths, $\tau_{\mbox{\tiny T}} \sim \tau_a$, naturally arise in
   the \Ha model on day 144 for the hydrogen ionization fraction of
   $0.04-0.05$ (see Appendix~\ref{secA1}) in line with the observed
   \Ha luminosity of $\approx4\times10^{40}$\ergs.
To summarize, the boxy asymmetric \Ha emission on day 144 is compatible with
   the spherical geometry of the CS interaction.

%===============================================================================
\subsection{Wind density and mass-loss rate}
\label{sec:winden}
%-------------------------------------------------------------------------------
%
\begin{figure}
   \includegraphics[width=\columnwidth, clip, trim=10 17 28 286]{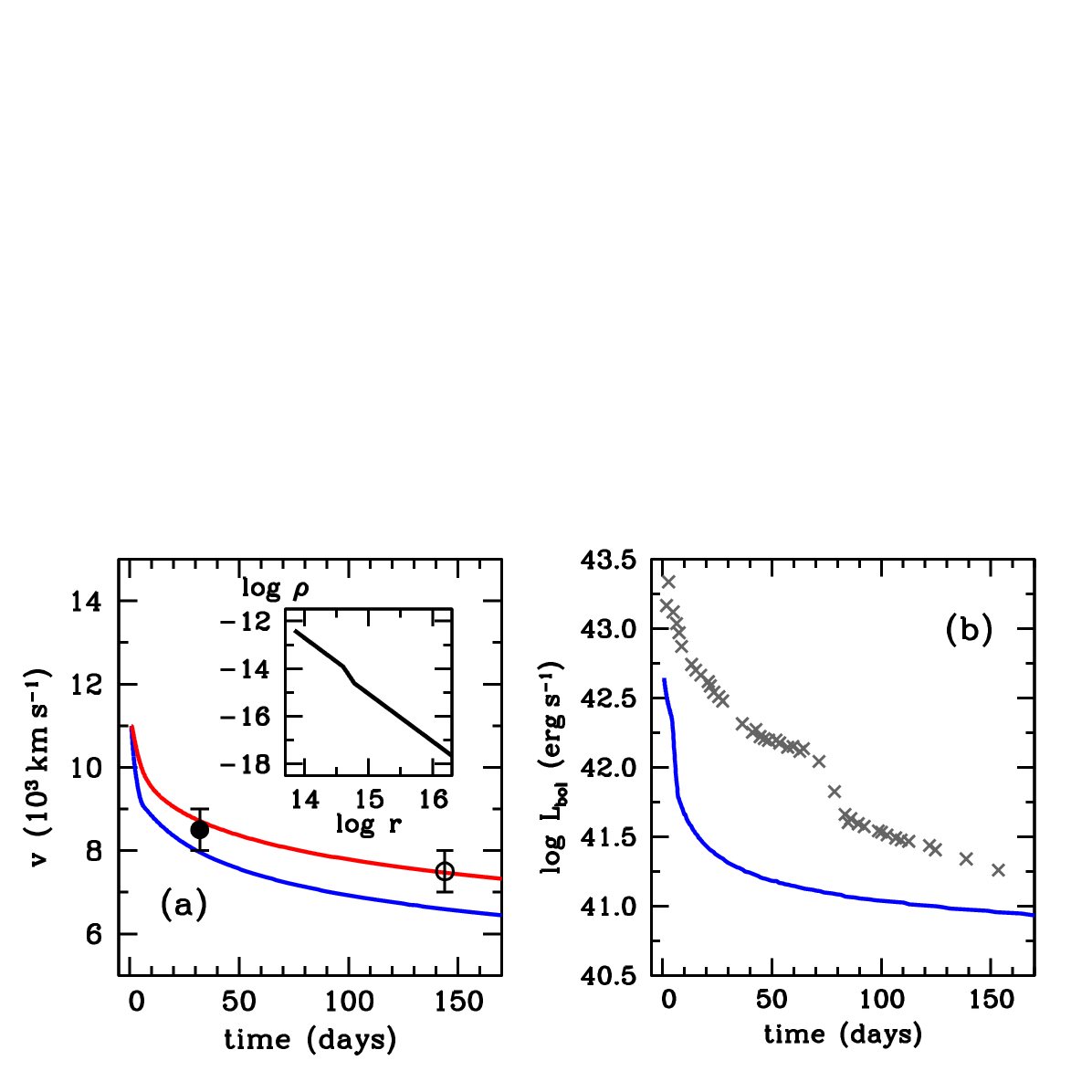}
   \caption{%
   Panel (a): expansion velocity of the CDS (\emph{blue line\/}) and
      the unshocked ejecta at the reverse shock (\emph{red line\/})
      in the model of the SN/CSM interaction compared to the observational
      estimates from \Hd absorption on day 33 and from \Ha emission on day 144.
   The inset shows the wind density.
   Panel (b): the combined luminosity of the forward and reverse shocks
      (\emph{blue line\/}) is significantly lower than the observed bolometric
      luminosity (black crosses).
   }
   \label{fig:wind}
\end{figure}
Given the spherical geometry, the ejecta deceleration can be used to recover
   the wind density parameter $w = \dot{M}/u$.
To this end we calculate the maximum velocity of the unshocked ejecta based on
   the thin-shell approximation \citep{Chevalier_1982b,  Chugai_2001} assuming
   the steady wind $\rho = (w/4\pi)r^{-2}$ and the SN density distribution
   $\rho = \rho_0/[1+(v/v_0)^9]$ that approximates the hydrodynamic model
   in the outer ejecta $v > 5000$\kms.

The ejecta deceleration is reproduced (Fig.~\ref{fig:wind}a) for the wind
   parameter $w = 1.1\times10^{16}$\gcm that implies the pre-SN mass-loss
   rate $\dot{M} = 6.6\times10^{-4}$\Msyr for the wind velocity of 35\kms
   reported by \citet{Andrews_2025}.
The DCS mass constrained by the deceleration down to the velocity of 8500\kms
   during the first month is $\approx 6\times10^{-3}$\Msun.
This value is twice as large compared to the DCS mass estimate assuming the 
   $\tau_{\mbox{\tiny T}} \sim 2$ (Section~\ref{sec:genphys}).  
Assuming the wind outflow velocity of 35\kms, the DCS has been formed
   during 6 years before the core collapse with the average mass-loss
   rate of $\sim$10$^{-3}$\Msyr.
The derived value is consistent with the range of $10^{-3} - 10^{-2}$\Msyr
   \citep{Andrews_2025} inferred from the early ($t < 7$ days) spectra being
   compared to the synthetic spectra computed earlier by \citet{Dessart_2017}.

The luminosity powered by the CS interaction does not affect significantly
   the bolometric luminosity through the observed epoch (Fig.~\ref{fig:wind}b).
At the late unobserved epoch, $t \gtrsim 250$ days, the CS interaction, likely,
   will dominate the SN luminosity powered by the $^{56}$Co radioactive decay.

%===============================================================================
\section{Discussion}
\label{sec:disc}
%-------------------------------------------------------------------------------
%
\begin{table}
\centering
\caption{Hydrodynamic models of type IIP supernovae}
\label{tab:sumtab}
\begin{tabular}{@{ } l  @{ } c  @{ } c @{ } c @{ } c @{ } c}
\toprule
\noalign{\smallskip}
 SN & $R_0$ & $M_{ej}$ & $E$ & $M_{\mathrm{Ni}}$ & $v_{\mathrm{Ni}}^{max}$ \\
    & (\Rsun) & (\Msun) & ($10^{51}$\,erg) & (\Msun) & (km\,s$^{-1}$) \\
\noalign{\smallskip}
\midrule
\noalign{\smallskip}
 1987A  &  35  & 18   & 1.5    & 0.0765 &  3000 \\
1999em  & 500  & 19   & 1.3    & 0.036  &  660  \\
2000cb  &  35  & 22.3 & 4.4    & 0.083  &  8400 \\
 2003Z  & 230  & 14   & 0.245  & 0.0063 &  535  \\
2004et  & 1500 & 22.9 & 2.3    & 0.068  &  1000 \\
2005cs  & 600  & 15.9 & 0.41   & 0.0082 &  610  \\
2008in  & 570  & 13.6 & 0.505  & 0.015  &  770  \\
2009kf  & 2000 & 28.1 & 21.5   & 0.40   &  7700 \\
2012A   &  715 & 13.1 & 0.525  & 0.0116 &  710  \\
2013ej  & 1500 & 26.1 & 1.4    & 0.039  &  6500 \\
 2016X  &  436 & 28.0 & 1.73   & 0.0295 &  4000 \\
2017gmr &  525 & 22.0 & 10.2   & 0.110  &  3300 \\
2018gj  &  775 & 23.4 & 1.84   & 0.031  &  5280 \\
2020jfo &  400 &  6.2 & 0.756  & 0.013  &  1600 \\
2024bch & 1250 & 13.0 & 2.0    & 0.075  &  7400 \\
\botrule
\end{tabular}
\end{table}
The primary goal has been to recover parameters of the short plateau type IIP
   SN~2024bch and to explore effects of the CS interaction.
We find the hydrodynamic pre-SN mass of 14.6\Msun, the explosion energy of
   $2\times10^{51}$\,erg, the pre-SN radius of 1250\Rsun, and the ejected
   $^{56}$Ni mass of 0.075\Msun.
The early escape of gamma-quanta indicates the $^{56}$Ni ejecta extension
   up to $\approx$7400\kms.
This value exceeds the maximal $^{56}$Ni velocity of $\approx$5700\kms
   produced by the 3D model B15-3 of SN~1987A with the explosion energy of
   $2.6\times10^{51}$\,erg \citep{Utrobin_2015}.

The modeling of the early stage of the CS interaction in the framework of the RHD
   reveals an interesting physics of the transition from the SBO to the 
   matter-dominated adiabatic shock with the gas temperature of $\sim$10$^9$\,K.
Such a modeling provides a promising diagnostic tool for the close CS environment
   based on the hard X-ray data.
It is noteworthy that we did not find another case of the similar RHD computation
   of the transition from the SBO in the exploding RSG till the adiabatic shock
   regime with the temperature of $\sim$10$^9$\,K.

We find that the radiative acceleration of the CSM predicts
   the preshock velocity of 500\kms on day 8, twice as small compared to
   the value of $\approx$1000\kms inferred from the \Ha absorption on day 8
   \citep{Andrews_2025}.
We argue that the X-rays from the forward shock are able to heat the preshock
   gas enough to provide a high expansion line opacity and therefore
   a more efficient radiative acceleration.
Next generation of hydrodynamic model could implement the X-ray emission and
   possibly resolve the problem of a high preshock velocity.
The additional acceleration by the CR precursor requires in this case
   the high DSA efficiency of $\approx$5\%, which seems to be doubtful
   on day 8.

The inferred wind density parameter of the pre-SN, $w = 1.1\times 10^{16}$\gcm,
   is larger by a factor of $10 - 10^2$ compared to the wind density recovered
   from the radio data for ordinary SNe~IIP \citep[cf.][]{Chevalier_2006}.
Remarkably, the wind density parameter of SN~2024bch is the same as that of
   the type IIL SN~1998S \citep{Chugai_2001}, although the DCS mass of SN~2024bch
   ($0.003-0.006$\Msun) is smaller by a factor of ten than that of SN~1998S.

The boxy component of the \Ha emission on day 144 with the estimated luminosity
   $L_b \approx 3\times10^{40}$\ergs is expected to be powered largely by
   the X-ray emission of both reverse and forward shocks.
Indeed, the CS interaction model predicts that the power of the X-ray emission
   injected in the SN ejecta including the CDS is $L_{inj} = 9\times10^{40}$\ergs
   that is three times larger than $L_b$.
Yet, the expected conversion factor of the absorbed power of the X-ray radiation
   to the \Ha luminosity is
   $\alpha_{32}E_{23}/(\alpha_{\mbox{\tiny B}}\mbox{Ry}) \approx 0.1$,
   three times smaller than the ratio $L_b/L_{inj}$.

The apparent energy deficit could be covered in two ways.
First, the wind clumpiness could increase the X-ray luminosity of the forward
   shock that powers the \Ha line.
Second, a significant fraction of the \Ha luminosity can be produced by
   the CDS material mixed with the hot gas of the forward shock due to
   the Rayleigh-Taylor instability.
The area of the contact surface between the hot forward shock gas
   ($T \approx 8\times10^8$\,K) and the cold CDS fragments is of the order of
   $S \sim 4\pi R_{cds}^2$.
Assuming the saturated regime of the electron conductivity with the heat flux
   $q_{sat} = 0.4(2kT/\pi m_e)^{0.5}n_ekT$ \citep{Cowie_1977},
   one expects the maximal injected power into the cold gas to be
   $L_{inj} = q_{sat}S \approx 1.3\times 10^{42}$\ergs.
Two percent of this power converted into the \Ha luminosity would be sufficient
   to eliminate the energy deficit.

The extended $^{56}$Ni mixing favors the early escape of gamma-quanta of the
   $^{56}$Co decay.
Originally this phenomenon was found by the SMM satellite for the type IIP
   SN~1987A with the detected flux of about $6.5\times10^{-4}$\pcms
   in the 847\,keV and 1238\,keV lines \citep{Matz_1988}.
To estimate the flux from a type IIP SN identical to SN~2024bch, but at
   the fiducial distance of 10\,Mpc, we consider the uniform ejecta of
   13\Msun with the energy of $2\times10^{51}$\,erg and the $^{56}$Ni mass
   of 0.07\Msun mixed uniformly.
The flux of escaping unscattered 847\,keV quanta at the optimal age of
   100 days in this case is $3\times10^{-7}$\pcms, nearly $2\times10^3$ times
   lower compared to SN~1987A.
Therefore, the detection of $^{56}$Co gamma-quanta from SN IIP even with
   the extended $^{56}$Ni ejecta currently is beyond reach.

\begin{figure}
   \includegraphics[width=\columnwidth, clip, trim=5 113 28 110]{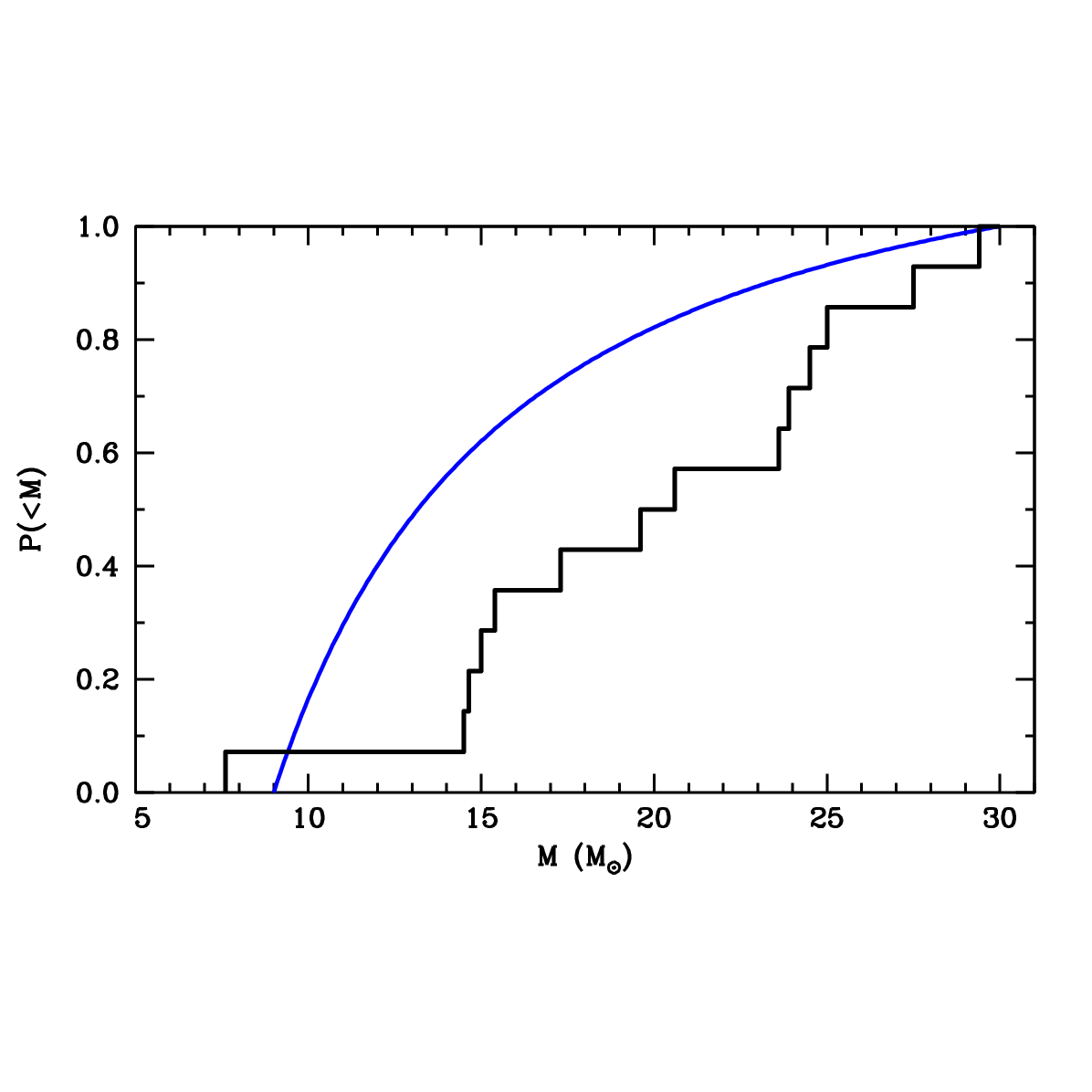}
   \caption{%
   Distribution function of pre-SN mass for the sample of 14 well-observed
      SNe~IIP studied hydrodynamically in a uniform approach (\emph{histogram\/})
      compared to the Salpeter IMF (\emph{blue line\/}).
   }
   \label{fig:salp}
\end{figure}
The overall mass distribution $P(<M)$ for the sample hydrodynamically explored
   SNe~IIP (Table \ref{tab:sumtab}) shows a pronounced dearth of stars with
   the mass of $<$15\Msun, when compared to the Salpeter IMF (Fig.~\ref{fig:salp}).
We rule out that the hydrodynamic modeling might systematically overproduce
   the ejecta mass.
Indeed, there is a consensus between different models for the well-studied
   SN~1999em \citep[cf.][]{Utrobin_2007, UWJM_2017} and
   SN~1987A \citep[cf.][]{Utrobin_2005, Utrobin_2021}.
The deficit of low-mass progenitors should have another explanation.

One can admit, e.g., that low-mass SNe~IIP are predominantly subluminous
   that is hinted by type IIP SN~2003Z and SN~2012A.
If this is the case, the sample of well-observed SNe~IIP that is used for
   the hydrodynamic modeling will be depleted in the low-mass range
   due to the observational selection.

Another possibility is that low-mass progenitors that have a lower binding
   energy loose their hydrogen envelope more easily compared to more massive
   progenitors.
In that case low-mass progenitors largely do not end up as SNe~IIP
   in line with the conjecture by \citet{Chevalier_2006}.

%===============================================================================
\section{Conclusions}
\label{sec:concl}
%-------------------------------------------------------------------------------
We conclude with the  major results:
\begin{itemize}
\item
The SN~2024bch is the explosion of a massive RSG with the pre-SN mass of
   $14-15$\Msun, the pre-SN radius of 1250\Rsun, the explosion energy of
   $2\times10^{51}$\,erg, and the radioactive $^{56}$Ni mass of 0.075\Msun. 
\item
The radioactive tail slope indicates the extension of $^{56}$Ni up to 7400\kms.
\item
The pre-SN was surrounded by the DCS with the mass of $0.003-0.006$\Msun
   within $6\times10^{14}$\,cm.
\item 
We estimate the wind density parameter $\dot{M}/u \approx 10^{16}$\gcmq
   that implies the mass-loss rate of $6\times10^{-4}$\Msyr.
\item 
The \Ha asymmetry of the broad boxy component on day 144 is due to
   the Thomson scattering and the Paschen continuum absorption and
   does not require the asphericity of the CS interaction.
\end{itemize}

\backmatter

\bmhead{Acknowledgements}
We thank Jen Andrews for the sharing of the \Ha spectrum.

\bmhead{Author contribution}
V.U. and N.C. contributed equally to this work.

\bmhead{Funding}
Not applicable.

\bmhead{Data availability}
No datasets were generated or analyzed during the current study.

\bmhead{Materials availability}
Not applicable.

\bmhead{Code availability} 
Not applicable.

\section*{Declarations}

\subsection*{Ethics approval and consent to participate}
Not applicable.

\subsection*{Consent for publication}
Not applicable.

\subsection*{Competing interests}
The authors declare no competing interests.

%===============================================================================
\begin{appendices}
%-------------------------------------------------------------------------------
\section{ }
\label{secA1}
%-------------------------------------------------------------------------------
Here we estimate the optical depth of the Paschen continuum
   in SN~2024bch on day 144.
The SN ejecta is set to be a uniform homologously expanding sphere with
   the mass $M = 13$\Msun, the hydrogen abundance X = 0.7, and the boundary
   velocity $v_0 = 7500$\kms at the age of 144 days.
The hydrogen number density at this moment is $n = 3.2\times10^8$\cmq.
For a given hydrogen ionization $x = n_e/n$ and the electron temperature
   $T_e = 6000$\,K (result is not sensitive to $T_e$), one can easily obtain
   the total \Ha luminosity and Thomson optical depth.
To infer the Paschen optical depth $\tau_a$, one need to solve the following
   balance equations for the hydrogen population on the third level $n_3$
\begin{eqnarray}
n_2(A_{21}\beta_{12} + q_{21}ne + A_{2q}) = \alpha_{\mbox{\tiny B}}n_e^2 \\
n_3A_{32}\beta_{23} = \alpha_{32}n_e^2.
\end{eqnarray}
Here $\beta_{ik} \approx 1/\tau_{ik}$ is the local escape probability for
   the line with the Sobolev optical depth $\tau_{ik} \gg 1$,
   $q_{21}$ is the collisional de-excitation coefficient,
   $A_{2q} = 2.06$ s$^{-1}$ is the two-photon decay rate,
   $\alpha_{\tiny B}$ is the Case B recombination coefficient,
   $\alpha_{32} = \alpha_{\mbox{\tiny B}} - \alpha_2$ is the effective
   \Ha recombination coefficient.
The inferred  population $n_3$ suggests the Paschen continuum optical depth
   at \Ha $\tau_a = \sigma_3(\lambda/\lambda_3)^{2.5}n_3v_0t$ with
   $\sigma_3 = 2.4\times10^{-17}$ cm$^2$, where $\lambda = 6563$\A, 
   $\lambda_3 = 8206$\A.
In Table~\ref{tab:paschen} we show the escaping \Ha luminosity (emitted
   \Ha luminosity multiplied by a factor of 0.7), the Thomson optical depth
   $\tau_{\mbox{\tiny T}}$ and the absorption optical depth $\tau_a$
   for two values of $x$.

\begin{table}
\centering 
\caption{ }
\label{tab:paschen}
\begin{tabular}{c c c c}
\toprule
\noalign{\smallskip}
x & $L$ ($10^{40}$\ergs) & $\tau_{\mbox{\tiny T}}$ & $\tau_a$ \\
\noalign{\smallskip}
\midrule
\noalign{\smallskip}
0.04     & 3.2  & 0.79  & 0.42  \\
0.05     & 5.3  & 0.99  & 0.95  \\
\botrule
\end{tabular}
\end{table}
\end{appendices}

%===============================================================================

\end{document}